\documentclass[pra,a4paper,aps,twocolumn,showpacs,superscriptaddress,groupedaddress]{revtex4}
\usepackage[dvips]{graphicx}
\usepackage{ae}
\usepackage[T1]{fontenc}
\usepackage[ansinew]{inputenc}
\usepackage{amsmath}
\usepackage{amssymb}
\usepackage{graphicx}
\usepackage{color}
\usepackage[colorlinks]{hyperref}
\usepackage{lscape}
\begin{document}
\author{Biswajit Paul}
\email{biswajitpaul4@gmail.com}
\affiliation{Department of Mathematics, South Malda College, Malda, West Bengal, India.}
\title{Revealing Hidden Genuine Tripartite Nonlocality}
\author{Kaushiki Mukherjee}
\email{kaushiki_mukherjee@rediffmail.com}
\affiliation{Department of Mathematics, Dr. A. P. J. Abdul Kalam Government College,Newtown, Kolkata, India.}
\author{Debasis Sarkar}
\email{dsappmath@caluniv.ac.in}
\affiliation{Department of Applied Mathematics, University of Calcutta, 92, A.P.C. Road, Kolkata-700009, India.}

\begin{abstract}
Nonlocal correlations arising from measurements on tripartite entangled states can be classified into two groups, one genuinely $3-$way nonlocal and other local with respect to some bipartition. Still, whether a genuinely tripartite entangled quantum state can exhibit genuine $3-$way nonlocality, remains a challenging problem so far as measurement context is concerned. Here we introduce a novel approach in this regard. We consider three tripartite quantum states none of which is genuinely $3-$way nonlocal in a specific Bell scenario (three parties, two measurements per party, two outcomes per measurement), but they can exhibit genuine $3-$way nonlocality when the initial states are subjected to stochastic local operations and classical communication (SLOCC). So, genuine $3-$way nonlocality is a resource, which can be revealed by using a sequence of measurements.

\end{abstract}
\date{\today}
\maketitle

\section{Introduction}
The seminal work of J. S. Bell refuted EPR argument \cite{Ein} claiming incompliances of Quantum theory. He in particular showed that there exist some correlations generated by measurements on a quantum system shared between distant parties that cannot be explained by any local hidden variable (LHV) theory \cite{Bel}. Such type of correlations, referred to as nonlocal correlations, are witnessed via violation of a Bell inequality \cite{Bru}. Apart from its importance as a foundational concept, nonlocality has also been used in various information-theoretic tasks \cite{Bur,Bae,Aci,Bad,Mck,Col,Pir}. For generation of nonlocal correlations, the quantum particles shared between distant parties must be entangled. However, the converse implication is not obvious. To be specific, though nonlocality can be considered as a generic notion for pure states \cite{Gin,Poe}, no such definite conclusion can be drawn for mixed states, as initially shown by Werner who presented a class of bipartite entangled states admitting a LHV model in the particular case of projective measurement \cite{Wer}. This model was later extended for general (positive-operator-valued-measurement, POVM)  measurements \cite{Bar} (see also \cite{Alm}). Such states
are referred to as local entangled states \cite{Aug}.\\
In this context, another important topic was discussed by Popescu\cite{Pop} and  Gisin \cite{Gis} who showed that some local entangled states, unable to produce nonlocal correlations under projective measurements, when subjected to suitable sequential measurements, can exhibit nonlocal behavior (violates the Bell-CHSH inequality \cite{Cla}). This process of revelation(or activation) of nonlocality of any state is referred to as its hidden nonlocality. In recent times it is shown that hidden nonlocality can be extracted even from those entangled states that admit a LHV model for POVMs \cite{Hir}. There exist some other related works in the literature showing revelation of nonlocality of local entangled states by performing joint measurements on several copies of the state \cite{Per,Mas,Lia,Nav,Pal,Cac,Brn}, or by placing many copies of the state in a quantum network \cite{Sen,Cav,Caa,Woj,Klo}. All of these works simply point out the fact that context of measurement is important to reveal nonlocality of quantum states and ongoing research activities in this direction imply that it is still a challenging field of research.\\
Though questions related to revelation of hidden nonlocality of local entangled states, have been extensively discussed for bipartite states, the relation between entanglement and hidden nonlocality for multipartite systems is almost unexplored so far. For multipartite scenario, one should intuitively expect some more interesting and novel phenomena, due to the complex structure of multipartite entanglement. In particular, there is a hierarchy of different notions of entanglement in tripartite systems, the strongest of them being genuine tripartite entanglement (GTE) \cite{Gun}. Analogous to entanglement in tripartite scenario, notion of genuine tripartite nonlocality (GTNL), discussed in \cite{Svt,Gal,Ban}, represents the strongest form of nonlocality for tripartite systems. \\
Now one may be interested to analyze whether hidden GTNL can be revealed under sequential measurements. In this context, Caban et al. \cite{Cab} gave an example of a class of tripartite pure states $\rho$ such that it does not violate the Svetlichny inequality \cite{Svt} whereas $\rho\bigotimes \rho$ can violate it and hence can exhibit Svetlichny's notion of GTNL. They however referred this phenomenon as activation of violation of Svetlichny  inequality. Recently a weaker (than Svetlichny's notion of GTNL) definition of GTNL has been introduced in \cite{Gal,Ban}, known as genuine $3-$way NS nonlocality ($NS_2$ nonlocality), which is better motivated both physically and from information theoretic view point. In this paper, we address the following question: consider some genuinely tripartite entangled states that do not exhibit $NS_2$ nonlocality individually in a specific Bell scenario (three parties, two measurements per party, two outcomes per measurement) and also in hidden sense, i.e., even after being subjected to known useful local filters \cite{Hic}. Is it then possible to find some sequential measurement protocol so that the final state resulting from the measurement protocol using these $NS_2$ local states, exhibits $NS_2$ nonlocality? We provide strong numerical evidence to this open problem. To be precise, we have framed a protocol based on sequential measurements which we refer to as sequential measurement protocol (SMP, see Fig.1). It involves three different tripartite quantum states. These three states, none of which was individually $NS_2$ nonlocal in the specific Bell scenario and not even after application of known useful local filters, when used in the SMP, generates a quantum state which is $NS_2$ nonlocal. However, as $NS_2$ nonlocality of the final state is revealed starting from $NS_2$ local initial states in the specific Bell scenario, so such revelation of hidden $NS_2$ nonlocality can be considered as revelation of weak hidden nonlocality. Moreover, the SMP can be used in principle even in the case when each of the states initially possessed by the parties has arbitrary amount of genuine entanglement.  \\
Rest of our paper is organized as follows. In Sec. II, we give a brief introduction to some concepts and results which we will use in later sections. We introduce the sequential measurement protocol in Sec. III together with the states used in the protocol to exhibit hidden GTNL. In Sec. IV we discuss our observations in the context of revealing hidden GTNL. Finally we conclude in Sec.V discussing various aspects of our findings along with scope of future research works.
\begin{figure}[htb]
\includegraphics[width=2.4in]{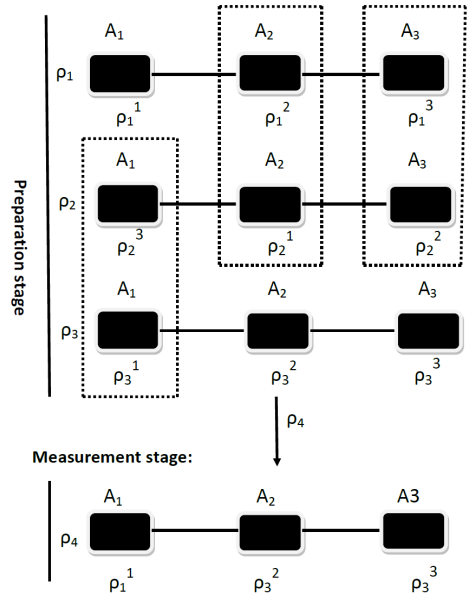}
\caption{\emph{The figure depicts a particular sequential measurement protocol involving three parties. $\rho_i^j$ denotes $j^{th}$ particle of $i^{th}$ state. Three states $\rho_1$ (Eq.(\ref{1})), $\rho_2$ (Eq.(\ref{2})) and $\rho_3$ (Eq.(\ref{3})) are distributed between the three parties $A_1$, $A_2$ and $A_3$ such that each of the three parties holds one particle from each of the three states. $A_1$ holds particles $\rho_1^1$, $\rho_2^3$ and $\rho_3^1$; $A_2$ holds $\rho_1^2$, $\rho_2^1$ and $\rho_3^2$ and $A_3$ holds particles $\rho_1^3$, $\rho_2^2$ and $\rho_3^3$. The sequential measurement protocol(SMP) is a particular example of WCCPI protocol. In the preparation stage, each of them performs Bell basis measurement on two of their respective three particles: $A_1$ performs Bell basis measurement on $\rho_2^3$ and $\rho_3^1$; $A_2$ performs Bell basis measurement on $\rho_1^2$, $\rho_2^1$ and $A_3$ performs Bell basis measurement on $\rho_1^3$ and $\rho_2^2$. Each bell-basis measurement is denoted by dotted box. Due to Bell basis measurements by each of the three parties and then  communication of the results among themselves, resultant state $\rho_4$ (Eq.(\ref{4})) is generated at the end of the preparation stage. $\rho_4$ (Eq.(\ref{4})) is shared between $A_1$, $A_2$ and $A_3.$ In the measurement stage, each of the parties $A_1$, $A_2$ and $A_3$ perform arbitrary projective measurements on their respective qubits of state $\rho_4$ (Eq.(\ref{4})). At the end of the measurement stage tripartite correlations will be generated in the SMP.
}}
\end{figure}
\section{PRELIMINARIES}
Before starting our discussion we provide all notions and facts necessary for further considerations.
\subsection{Genuine tripartite nonlocality}
 To analyze the nature of correlations shared between three systems, different forms of nonlocality can be considered. The local tripartite correlations have the form:
\begin{equation}\label{4i}
    P(a,b,c|x,y,z)=\sum_{\lambda}q_{\lambda}P_{\lambda}(a|x)P_{\lambda}(b|y)P_{\lambda}(c|z),
\end{equation}
where $a,\,b,\,c\,\in\{0,1\}$ denote the outputs and $x,\,y,\,z\,\in\{0,1\}$ denote inputs of the parties Alice, Bob and Charlie respectively, $0\leq q_{\lambda}\leq 1$ and $\sum_{\lambda}q_{\lambda}=1.$ $P_{\lambda}(a|x)$ is the conditional probability of getting outcome $a$ when the measurement setting is $x$ and $\lambda$ is the hidden state; $P_{\lambda}(b|y)$ and $P_{\lambda}(c|z)$ are similarly defined. Tripartite correlations that cannot be written as in Eq.(\ref{4i}) are called nonlocal. Bell type inequalities based on local realism (Eq.(\ref{4i})) fail to distinguish between bipartite and tripartite nonlocality \cite{Coi,Cer,Mer}. In order to detect GTNL, Svetlichny introduced  hybrid local-nonlocal form of correlations \cite{Svt}:
$$P(abc|xyz) =\sum_{\lambda}q_{\lambda}P_{\lambda}(ab|xy ) P_{\lambda}(c|z)+$$
\begin{equation}\label{1i}
    \sum_{\mu} q_{\mu}P_{\mu}(ac|xz) P_{\mu}(b|y ) +\sum_{\nu} q_{\nu}P_{\nu}(bc|yz) P_{\nu}(a|x);
\end{equation}
where $0\leq q_{\lambda},\,q_{\mu},q_{\nu}\,\leq1$ and $\sum_{\lambda}q_{\lambda}+\sum_{\mu}q_{\mu}+\sum_{\nu}q_{\nu}=1.$
This form of correlations are referred as Svetlichny local ($S_2$ local), otherwise Svetlichny nonlocal ($S_2$ nonlocal) \cite{Ban}. Based on this, Svetlichny designed a tripartite Bell type inequality (known as Svetlichny inequality):
\begin{equation}\label{2i}
 S\leq4.
\end{equation}
where $ S\,=\, \langle x_0y_0z_0\rangle+\langle x_1y_0z_0\rangle-\langle x_0y_1z_0\rangle +\langle x_1y_1z_0 \rangle$ $$  +\langle x_0y_0z_1\rangle-\langle x_1y_0z_1\rangle+\langle x_0y_1z_1\rangle +\langle x_1y_1z_1\rangle. $$
Violation of this inequality guarantees $S_2$ nonlocality, sufficient to detect GTNL. While Svetlichny's notion of GTNL is often referred to in the literature, it has certain drawbacks. As has been pointed out in \cite{Gal,Ban,Kau}, Svetlichny's notion of GTNL is so general that correlations capable of two-way signaling are allowed among some parties. This may lead to grandfather-style paradoxes \cite{Ban} and provide inconsistency in operational purposes \cite{Gal,Gae}. To remove this ambiguity, Bancal et al.\cite{Ban}, introduced genuine $3-$way NS nonlocality ($NS_2$ nonlocality). Suppose $P(abc|xyz)$ be the tripartite correlation satisfying Eq.(\ref{1i}) with non-signalling criteria imposed on the bipartite correlations terms,
\begin{equation}\label{4ii}
   P_{\lambda}(a|x)=\sum_b P_{\lambda}(ab|xy )\, ~~\forall\, a,\, x,\, y,
\end{equation}
\begin{equation}\label{4iii}
  P_{\lambda}(b|y)=\sum_a P_{\lambda}(ab|xy ) \, ~~\forall \,b, x,\, y.
\end{equation}
and similarly for $P_{\mu}(ac|xz)$ and $P_{\nu}(bc|yz).$
This form of correlations are called $NS_2$ local. Otherwise, they are $NS_2$ nonlocal. In analyzing the procedure of revelation of hidden  GTNL, we have used the necessary and sufficient criteria for detecting GTNL provided by the whole set of $185$ facet inequalities of the $NS_2$ local polytope in the presence of binary input and output (see Supplementary Material of \cite{Ban}). Svetlichny inequality constitutes the $185$th facet inequality. Throughout the paper we have used  projective measurements to check nature of correlations generated by some tripartite quantum states.\\
\subsection{Wirings And Classical Communication Prior To The Inputs(WCCPI Protocol)}
This protocol may be considered as a set of allowed operations that cannot create nonlocality i.e., interpret nonlocality as a resource, analogous to entanglement which cannot be created by Local Operations and Classical communication(LOCC). This type of protocol was first used in \cite{Gal} for framing multipartite nonlocality as a resource. The protocol introduced there  involved single measurement. Later it was extended for sequential measurements in \cite{Gag}. A \textit{valid} WCCPI protocol for sequential measurements\cite{Gag}, characterizing basically correlation terms generated in any sequential scenario, mainly consists of two stages: \textit{preparation stage} and \textit{measurement stage}. In the preparation stage the parties are allowed to perform measurements on their respective physical systems and then communicate the corresponding outputs among each other. As the parties have not yet received any input for the final Bell test(going to take place in the measurement stage), classical communication is allowed in the preparation stage. However, this communication cannot be used to generate any sort of nonlocal correlations. The inputs of the parties for the final stage, i.e., the measurement stage depend on outputs that are obtained  and communicated in the preparation stage. In the measurement stage no further communication is allowed between the parties. The permissible local operations of each party consist of processing the classical inputs and outputs and are referred to as \textit{wirings}. The sequential correlations generated at the end of the measurement stage help in characterizing nonlocality as a resource. As already discussed before, nonlocality cannot be created by WCCPI. So GTNL cannot also be created by WCCPI protocol. In our present topic of discussion, we have introduced a measurement protocol which may be considered as a WCCPI protocol.
\subsection{Genuine multipartite concurrence ($C_{GM})$}
We briefly now describe $C_{GM}$, a measure of genuine multipartite entanglement. For pure $n$-partite states($|\psi\rangle$), this measure defined as \cite{Zma} :
$C_{GM}(|\psi\rangle):= \textmd{min}_j\sqrt{2 (1-\Pi_j(|\psi\rangle))}$ where $\Pi_j(|\psi\rangle)$ is the purity of $j^{th}$ bipartition of $|\psi\rangle$. The expression of $C_{GM}$ for $X$ states are given in \cite{Has}. For tripartite $X$ states,
\begin{equation}\label{4v}
  C_{GM}=2\,\textmd{max}_i\{0,|\gamma_i|-w_i\}
\end{equation}
with $w_i=\sum_{j\neq i}\sqrt{a_jb_j}$ where $a_j$, $b_j$ and $\gamma_j(j=1,2,3,4)$ are the elements of the density matrix of tripartite X state:

 \[\begin{bmatrix}
       a_1 & 0    & 0    & 0   &  0   &  0    & 0    & \gamma_1 \\
       0   & a_2  & 0    & 0   &  0   &  0    & \gamma_2  &  0  \\
       0   & 0    & a_3  & 0   &  0   &  \gamma_3  & 0    &  0  \\
       0   & 0    & 0    & a_4 &  \gamma_4 &  0    & 0    &  0  \\
       0   & 0    & 0   &{\gamma_4}^\ast &b_4 & 0  & 0   &    0   \\
       0   & 0    &{\gamma_3}^\ast   &0 &  & b_3  & 0   &    0  \\
       0   &{\gamma_2}^\ast    & 0  & 0 & 0 & 0 & b_2 & 0 \\
       {\gamma_1}^\ast & 0 &0 & 0 & 0 & 0 & 0 & b_1 \\
 \end{bmatrix}\]
\section{Sequential Measurement Protocol}
Consider a measurement protocol connecting three distant observers $A_i(i=1,2,3)$. $n$ tripartite quantum states $\rho_j(j=1,2,...,n)$ can be used in the protocol. Let each of $n$ states $\rho_j(j=1,2,...,n)$ fails to reveal GTNL in the specific Bell scenario. Each of these $n$ states $\rho_j(j=1,2,...,n)$ can be distributed between the three parties $A_i(i=1,2,3)$ with  some specification in distribution of qubits among the parties such that each of the three parties holds one particle from each of the $n$ states.  So each of the parties holds $n$ qubits in his lab. This protocol is a particular example of WCCPI protocol. In the preparation stage, each party can perform some joint measurement on their respective $n-1$ particles and then communicate the results between themselves. At the end of measurements by all the three parties, $\rho_{n+1}$, a tripartite quantum state shared between $A_1,\, A_2$ and $A_3$, is generated. Clearly, as in any WCCPI protocol, the state $\rho_{n+1}$ is output specific, i.e., depends on the output of the joint measurements performed by the parties in the preparation stage. In the measurement stage of the protocol, each of the three parties can now perform arbitrary projective measurements on their share of the physical system $\rho_{n+1}$ but are not allowed to communicate among themselves thereby generating tripartite correlation terms whose nature can now be tested using some tripartite Bell inequality.  We refer to this protocol of sequential measurements by the three parties sharing $n$ states as Sequential Measurement Protocol (SMP).  Now we have already discussed before that GTNL cannot be created by WCCPI protocol. Hence generation of GTNL by SMP, starting from three local initial states, guarantee revelation of hidden GTNL by our SMP. Our SMP can be considered as a particular type of sequential measurement protocol via which hidden GTNL can be revealed, analogous to the sequential measurement protocol introduced by Popescu for revealing hidden bipartite nonlocality \cite{Pop}. We provide an explicit example of revelation of hidden GTNL for $n=3$ by using our SMP(see Fig.1). Suppose the three initial states shared between the three parties be given by:
\begin{equation}\label{1}
    \rho_1 = p_1 |\psi_f\rangle\langle \psi_f|+(1-
    p_1)|001\rangle\langle001|
\end{equation}
 with $|\psi_f\rangle=\cos\theta_1|000\rangle+\sin\theta_1|111\rangle$, $0\leq\theta_1\leq \frac{\pi}{4}$ and $0\leq p_1\leq 1$;
\begin{equation}\label{2}
   \rho_2=  p_2 |\psi_m\rangle\langle \psi_m|+(1-p_2)|010\rangle\langle010|
\end{equation}
with $|\psi_m\rangle=\frac{|000\rangle+|111\rangle}{\sqrt{2}}$ and $0\leq p_2\leq 1$;
\begin{equation}\label{3}
  \rho_3=  p_3 |\psi_l\rangle\langle \psi_l|+(1-p_3)|100\rangle\langle100|
\end{equation}
with $|\psi_l\rangle=\sin\theta_3|000\rangle+\cos\theta_3|111\rangle$, $0\leq\theta_3\leq \frac{\pi}{4}$ and $0\leq p_3\leq 1$. The $i$-th particle of each of $\rho_1(\rho_1^i)$ (Eq.(\ref{1})) and $\rho_3(\rho_3^i)$ is with the party $A_i(i=1,2,3)$ whereas the three particles of $\rho_2$, i.e., $\rho_2^1$, $\rho_2^2$ and $\rho_2^3$ are with parties $A_2$, $A_3$ and $A_1$ respectively. Hence each of the three parties $A_i(i=1,2,3)$ has three particles. Now in the preparation stage of the SMP, each of the three parties $A_i(i=1,2,3)$  performs Bell basis measurements on two of the three particles that each of them holds: $A_1$ performs Bell basis measurement on $3^{rd}$ particle of $\rho_2$($\rho_2^3$) and $1^{st}$ particle of $\rho_3$($ \rho_3^1$); $A_2$ performs Bell basis measurement on $2^{nd}$ particle of $\rho_1$($\rho_1^2$) and $1^{st}$ particle of $\rho_2$($ \rho_2^1$); $A_3$ performs Bell basis measurement on $3^{rd}$ particle of $\rho_1$($\rho_1^3$) and $2^{nd}$ particle of $\rho_2$($ \rho_2^2$). After all the three parties have performed Bell basis measurement on their respective particles, they communicate the results among themselves, as a result of which $\rho_4$ is generated at the end of the preparation stage. If the output of each of the measurement is $|\psi^{\pm}\rangle(\frac{|01\rangle\pm|10\rangle}{\sqrt{2}})$, the resultant state (correcting phase term) is given by:
\begin{equation}\label{4}
  \rho_4 = \frac{p_3 |\phi\rangle\langle\phi| + (1 - p_3)\sin^2 \theta_1|100\rangle\langle100|}{\sin^2 \theta_1+ p_3 \cos2\theta_1 \sin^2 \theta_3 }
\end{equation}
where $|\phi\rangle = \cos\theta_1 \sin\theta_3|000\rangle + \sin\theta_1 \cos\theta_3 |111\rangle$. Eq.(\ref{4}) points out that $\rho_4$ is independent of $p_1$ and $p_2$. Clearly the final state $\rho_4$ is obtained from the initial states $\rho_i (i=1,2,3)$ by means of post-selecting on particular results ($|\psi^{\pm}\rangle$) of local measurements. So preparation stage of this protocol can be considered as a particular instance of Stochastic Local Operations And Classical Communication (SLOCC). After $\rho_4$ is generated and shared between the parties in the preparation stage, each of the three parties $A_1$, $A_2$ and $A_3$ performs projective measurement on the state $\rho_4$ in the measurement stage. Now if the correlations generated from $\rho_4$ exhibits GTNL under the context that the initial  states $\rho_i(i=1,2,3)$ fail to reveal the same, then that guarantees generation of hidden GTNL in the SMP. However $\rho_4$ can be generated for some other specification of SMP protocol also, specially for some different arrangement of particles between the parties $A_i(1,2,3)$ and for different outputs of Bell measurements. Having designed the SMP, we are now going to present our results.\\
\section{Revelation of hidden genuine tripartite nonlocality}
In this section we discuss in details our observations which guarantee that the SMP introduced in the last section helps in revealing hidden GTNL. For that we consider two different notions of hidden GTNL: hidden $S_2$ nonlocality and hidden $NS_2$ nonlocality. Firstly we consider the former notion.
\subsection{Revelation of hidden Svetlichny nonlocality}
Existence of hidden $S_2$ nonlocality will be guaranteed if we can transform $S_2$ local $\rho_i(i=1,2,3)$ to $\rho_4$, capable of violating Eq.(\ref{2i}). Below we will show that the final state $\rho_4$, resulting from the preparation stage of the SMP, exhibits $S_2$ nonlocality, though the initial states $\rho_i(i=1,2,3)$ are $S_2$ local. The maximum value of the Svetlichny operator($S$) upto projective measurements, for state $\rho_i(i=1,2,3)$ is given by (see Appendix A) :
 $$B_1 = \max[4\sqrt{2}\, p_1\sin2\theta_1, 4|(1 - p_1 - p_1 \cos2\theta_1 )|],$$
  $$B_2 = \max[4\sqrt{2}\, p_2, 4(1 - p_2)]\,\,\,\,$$ and
 \begin{equation}\label{6i}
B_3 = \max[4\sqrt{2}\, p_3\sin2\theta_3, 4|(1 - p_3 - p_3 \cos2\theta_3 )|]
\end{equation}
respectively whereas that for the final state $\rho_4$, it is given by:
$$B_4 = \max[\frac{2\sqrt{2}\,p_3 \sin2\theta_1\sin2\theta_3}{\sin^2 \theta_1+ p_3 \cos2\theta_1 \sin^2 \theta_3},$$
\begin{equation}\label{6ii}
\frac{2|(1 - 2 p_3 \sin^2 \theta_3- \cos\theta_1 )|}{\sin^2 \theta_1+ p_3 \cos2\theta_1 \sin^2 \theta_3}].
\end{equation}
Since both the initial ($\rho_i,i=1,2,3$) and final states ($\rho_4$) belong to the class of tripartite X states, their amount of genuine entanglement can be measured by Eq.(\ref{4v}). For the initial states $\rho_i (i=1,2,3)$, the amount of GTE are given by:
 $$ C^{\rho_1}_{GM} = p_1 \sin 2\theta_1 ,$$
 $$ C^{\rho_2}_{GM} = p_2 $$ and
   \begin{equation}\label{6iii}
  C^{\rho_3}_{GM} = p_3 \sin 2\theta_3
   \end{equation}
   whereas that for $\rho_4$ is given by:
   \begin{equation}\label{6iv}
    C^{\rho_4}_{GM} = \frac{p_3 \sin2\theta_1\sin2\theta_3 }{2(\sin^2 \theta_1+ p_3 \cos2\theta_1 \sin^2 \theta_3)}.
   \end{equation}
The initial states $\rho_i(i=1,2,3)$ are  genuinely entangled for any nonzero value of the state parameters (Eq.(\ref{6iii})). It is clear from the maximum  value of Svetlichny operator(Eqs.(\ref{6i}), (\ref{6ii}))  and the measure of entanglement (Eqs.(\ref{6iii}), (\ref{6iv})) of both initial states and final state, that each of them is $S_2$ local for $C^{\rho_i}_{GM}\leq \frac{1}{\sqrt{2}}(i=1,2,3,4).$ Thus existence of hidden $S_2$ nonlocality can be observed
 if for some fixed values of the parameters of the three initial $S_2$ local states ($C^{\rho_i}_{GM}\leq\frac{1}{\sqrt{2}}$), the final state can have $C^{\rho_4}_{GM}>\frac{1}{\sqrt{2}}$. Now for $\theta_1=0.1$, $p_2\leq\frac{1}{\sqrt{2}}$, $\theta_3=0.144$ and $p_1\,,p_3\in[0,1],$ each of the initial states is $S_2$ local ($C^{\rho_i}_{GM}\leq\frac{1}{\sqrt{2}}$) whereas the resultant state $\rho_4$ violates Svetlichny  inequality ($C^{\rho_4}_{GM}>\frac{1}{\sqrt{2}}$) for $p_3 \geq 0.5055$. In this explicit example, initial genuinely tripartite entangled states do not violate Svetlichny inequality but when used in preparation stage of our SMP, they can generate a state which exhibits $S_2$ nonlocality. This guarantees existence of hidden $S_2$ nonlocality for $p_3 \in [0.5055,1]$ (See Fig.2). \\
Now use of local filters is known to be a standard method to reveal hidden nonlocality. Interestingly, our SMP can reveal hidden $S_2$ nonlocality using some initial states which are even incapable of exhibiting hidden $S_2$  nonlocality(i.e., cannot reveal $S_2$  nonlocality after being subjected to known useful local filters \cite{Hic}). We proceed forward with an example. Let known useful local filters be applied on each of the three initial states $\rho_i(i=1,2,3)$ to reveal hidden $S_2$ nonlocality of the individual state. The maximum value of Svetlichny operator $S$ (Eq.(\ref{2i})), in terms of state parameters, for each of the three states $\rho_i(i=1,2,3)$, after applying known useful local filters, are derived(see Appendix B). Maximum values of $S$, in turn, provide constraints on the state parameters such that each of initial states $\rho_i$, has no $S_2$ nonlocality even after being subjected to local filtering. For a particular instance, when $\theta_1 =0.1$, $\rho_1$, after being filtered, remains still $S_2$ local  for  $p_1\in[0, 0.5025]$. Similarly second state($\rho_2$), after being subjected to filtering remains $S_2$ local for $p_2\in[0,0.6666]$, but the range of $p_3$ for which $\rho_3$ exhibits $S_2$ nonlocality remains unaltered both before and after filtering when $\theta_3 = 0.144$ (see Appendix B). Hence each of the initial states $\rho_i (i=1, 2, 3)$, under some restricted range of state parameters, has no hidden $S_2$ nonlocality. Now if these initial states under the said restricted range are used in the initial stage(preparation stage) of our SMP then $S_2$ nonlocality will be revealed for $p_3\in [0.5055,1]$. However, this range of revelation of hidden $S_2$ nonlocality in our SMP remains the same when the states $\rho_i(i=1,2,3)$ are used without being filtered. This example thus suffices to justify our claim that our SMP can reveal hidden $S_2$ nonlocality even from some initial states which have no hidden $S_2$ nonlocality. This in turn points out the utility of SMP over the standard procedure of using local filters for revelation of hidden $S_2$ nonlocality.
 In the context of our discussion, it should be pointed out that in \cite{Cab}, hidden $S_2$ nonlocality was observed. But our method and the results differ from that discussed in \cite{Cab}. It was shown there that if the three parties share two identical copies of the genuinely entangled state $\kappa$ such that each of $\kappa$ does not violate Svetlichny inequality, then $\kappa\bigotimes\kappa$ can violate Svetlichny inequality, maximal amount of violation being 4.2418. Moreover in our  SMP, there exist initial states $\rho_i(i=1,2,3)$ which do not violate Svetlichny inequality whereas the final state $\rho_4$ generated from the initial stage(preparation stage) of our SMP(Fig.1) can violate Svetlichny inequality maximally. For instance, if we consider $\rho_i(i=1,2,3)$ as the three initial states with $\theta_3=\theta_1$ and $p_3=1$, then with $S_2$ local version of these initial states, i.e. under some restricted range of $\theta_1$, $p_1$ and $p_2$(Eq.(\ref{6i})): $0 < \sin 2\theta_1 \leq \frac{1}{\sqrt{2}}$, $0< p_1 \leq 1$, and $0< p_2 \leq \frac{1}{\sqrt{2}}$, maximally entangled state $|\psi_m\rangle$ is obtained. Even with arbitrarily lower values of $\theta_1$, $p_1$ and $p_2$, i.e., with initial states having lower values of $C_{GM}$, $|\psi_m\rangle$ can be obtained and hence maximal violation of Svetlichny inequality can be observed. This in turn points out utility of our SMP to check the existence of hidden $S_2$ nonlocality from experimental perspectives.\\
\subsection{Revelation of hidden genuine $3-$way NS nonlocality}
Initial states used so far were $S_2$ local. However some of them were genuinely $3-$way NS nonlocal as they can violate one of the $185$ facets (except Svetlichny inequality). So revelation of hidden $S_2$ nonlocality via violation of Svetlichny inequality does not guarantee existence of hidden $NS_2$ nonlocality. For that purpose, all the initial states must be $NS_2$ local and the final state(resulting from the preparation stage of the SMP) must violate atleast one of these $185$ facets. We now proceed to present instances in support of our claim that hidden $NS_2$ nonlocality can be revealed by our SMP.   \\
\begin{figure}[htb]
\includegraphics[width=2.4in]{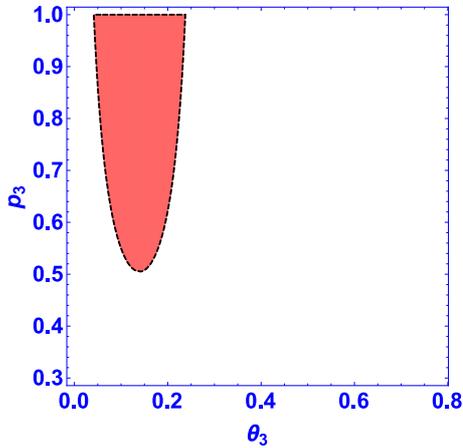}
\caption{\emph{The region of revelation of hidden $S_2$ nonlocality for $\theta_1=0.1$ where the initial states $\rho_i (i=1,2,3)$ are $S_2$ local. Similar type of regions of revelation of hidden $S_2$ nonlocality can be observed for different values of $\theta_1.$
}}
\end{figure}
Consider three genuinely tripartite entangled states($\rho_i,i=1,2,3$) satisfying all the $185$ facets of $NS_2$ local polytope (for some restricted range of state  parameters). Precisely, each of the three states are $NS_2$ local. If the final state $\rho_4$, resulting from the preparation stage of SMP, violates at least one of the $185$ facets, then that implies revelation of hidden $NS_2$ nonlocality. For instance, consider $\rho_1$ (Eq.(\ref{1}))  with $\theta_1=0.1,$ $p1<0.509$, $\rho_2$ (Eq.(\ref{2})) with $p_2<0.6672$ and $\rho_3$ (Eq.(\ref{3})) with $\theta_3=0.3$, $p_3<0.9198$, then these initial states satisfy all the $185$ facet inequalities. The final state $\rho_4$(Eq.(\ref{4}))where $\theta_1 = 0.1$, $\theta_3 = 0.3$, violates some of the facet inequalities over varying range of the state parameter $p_3$. For $p_3\geq 0.105$, $16$th facet inequality (same numbering as in \cite{Ban} has been used for convenience) is violated. This implies existence of hidden $NS_2$ nonlocality in the range  $p_3\in[0.105,0.9198]$. These ranges of $p_1$, $p_2$, $p_3$  are found by numerical optimization using Mathematica software \cite{Wol}(see Appendix A). There exist many other specific $NS_2$ local initial states belonging to the three families of tripartite mixed states(Eqs.(\ref{1}), (\ref{2}), (\ref{3})) for which the state generated by the preparation stage of our  SMP(Fig.1) can reveal hidden $NS_2$ nonlocality. We have thus succeeded to show the existence of hidden $NS_2$ nonlocality by our SMP. Some numerical observations are enlisted  in Table(\ref{table1}). These observations justify our claim that arbitrarily small amount of GTE suffices to reveal hidden $NS_2$ nonlocality. Analogous to our approach in the case of $S_2$ nonlocality, here we consider three initial states, none of which is $NS_2$ nonlocal even after being subjected to filtering. Then these states when used in our SMP generate $NS_2$ nonlocal correlations. We provide with an explicit illustration in support of our claim. Let known useful filters be applied on each of the three initial states $\rho_i(i=1,2,3)$. Fixing the state parameter of $\rho_1$ to be $\theta_1=0.1$, we apply known useful filters over it. After being filtered, it remains $NS_2$ local for $p_1\in [0, 0.5025]$. Similarly second state($\rho_2$), after filtration remains $NS_2$ local for $p_2\in[0,0.6666]$. However, for $\theta_3 = 0.3$, the range of $p_3$ for which $\rho_3$ exhibits $NS_2$ nonlocality does not change after applying filtering  operation (see Appendix B). So for each of the three initial states $\rho_i(i=1,2,3)$, after being subjected to local filtering, there exist some range of state parameters for which $NS_2$ nonlocality cannot be revealed. If these initial states under the said restricted range are used in our SMP then $NS_2$ nonlocality will be revealed for $p_3\in [0.105,0.9198]$. However, analogous to revelation of hidden $S_2$ nonlocality, this range of revelation of hidden $NS_2$ nonlocality in our SMP remains the same when the states $\rho_i(i=1,2,3)$ are used without being subjected to filtration. Thus our SMP turns out to be more efficient compared to the standard procedure of using local filters for revelation of hidden $NS_2$ nonlocality.
 \begin{table}[htp]
\begin{tabular}{|c|c|c|c|}
			\hline
			$\rho_1$ & $\rho_2$ &$\rho_3$&Revelation Range\\
			\hline
			$p_1$$<$$0.509$ &  $p_2$$<$$0.6672$&$\theta_3$$=$$0.1$,\, $p_3$$<$$0.9901$&$p_3$$\in$$[0.504,0.9901]$\\
			\hline
			$p_1$$<$$0.509$ &  $p_2$$<$$0.6672$&$\theta_3$$=$$0.5$,\, $p_3$$<$$0.8135$&$p_3$$\in$$[0.0425,0.8135]$\\
			\hline
			$p_1$$<$$0.509$ &  $p_2$$<$$0.6672$&$\theta_3$$=$$0.7$,\, $p_3$$<$$0.7072$&$p_3$$\in$$[0.0243,0.7072]$\\
			\hline
			$p_1$$<$$0.509$ &  $p_2$$<$$0.6672$&$\theta_3$$=$$0.785$,\, $p_3$$<$$0.6677$&$p_3$$\in$$[0.0202,0.6677]$\\
			\hline
\end{tabular}\\
\caption{The range of revelation of hidden genuine $3-$way NS nonlocality for state parameter $p_3$ is given in the table for different fixed values of the state parameters of the $NS_2$ local initial $\rho_i(i=1,2,3)$. These values were found by numerical optimization(by Mathematica Software). Here we consider a fixed value of state parameter($\theta_1$) of $\rho_1$, $\theta_1$$=$$0.1$. Clearly range of revelation varies with variation of $\theta_1$.}
\label{table1}
\end{table}

\section{Discussion}
From our discussion so far we conclude that genuine $3-$way NS nonlocality is some kind of resource, which can be revealed by a sequence of measurements. Usually it is believed that standard Bell scenario(i.e., in each run of the experiment, non-sequential local measurements are performed on a single copy of an entangled state) is suitable for a quantum state to exhibit genuine $3-$way NS nonlocality. Our present work, however can be considered as an approach deviated from this usual belief. We have shown that three tripartite quantum states, unable to reveal genuine $3-$way NS nonlocality in the standard Bell scenario, when used in our Sequential Measurement Protocol (SMP) can generate a state which is genuinely $3-$way NS nonlocal. This implies that hidden genuine $3-$way NS nonlocality can be revealed. Even our SMP emerges to be more efficient to reveal hidden genuine $3-$way NS nonlocality compared to the standard procedure of using known useful local filters.\\ Besides, the preparation stage of our SMP protocol can also be interpreted as an entanglement swapping protocol. Consequently via this protocol we can give an affirmative answer for tripartite system to a query posed by Sen et al. \cite{Sen}: consider some local entangled states, is it possible to find some entanglement swapping process, so that the swapped states, resulting from it, are capable of showing nonlocal behavior?\\  There are a number of possible generalizations of the above results. One may explore whether for any genuinely tripartite mixed entangled state, existence of at least one suitable SMP is guaranteed under which revelation of hidden GTNL is possible. Also, it will be interesting to generalize our SMP so as to demonstrate $n$ partite hidden genuine nonlocality. Moreover, till now there exist various experimental works demonstrating tripartite nonlocality \cite{Lav,Ham,Wag} and also hidden bipartite nonlocality \cite{Kwa}. In that context, one may expect to use our protocol for experimental verification of hidden GTNL. Besides, as GTNL implies GTE, our SMP can be used in the laboratory to detect GTE of the initial states in a device independent way \cite{Baa,Bai}.\\
\textit{Acknowledgments:}
The authors acknowledge useful suggestions from anonymous referees which have helped not only to motivate our paper a lot but also helped us to present our findings in a more compact form. The authors also acknowledge fruitful discussions with A.Sen and S.Karmakar. The author D.S. acknowledges DST-SERB for financial support.

\section{Appendix}
\subsection{Checking facets of $NS_2$ local polytope}
\textit{Derivation of $B_i(i=1,2,3,4)$:}
In order to obtain the maximum value $B_1$ (Eq.(\ref{6i}) of main text) of the Svetlichny operator $S$ (Eq.(\ref{2i})) upto projective measurements  we follow the method used in \cite{Kau}. We consider the following measurements: $x_0 = \vec{x}.\vec{\sigma_1} $ or $x_1 = \vec{\acute{x}}.\vec{\sigma_1}$ on $1^{st}$ qubit, $y_0 = \vec{y}.\vec{\sigma_2} $ or $y_1 = \vec{\acute{y}}.\vec{\sigma_2}$ on $2^{nd}$ qubit, and $z_0 = \vec{z}.\vec{\sigma_3} $ or $z_1 = \vec{\acute{z}}.\vec{\sigma_3}$ on $3^{rd}$ qubit, where $\vec{x},\vec{\acute{x}},\vec{y},\vec{\acute{y}}$ and $\vec{z},\vec{\acute{z}}$ are unit vectors and $\sigma_i$ are the spin projection operators that can be written in terms of the Pauli matrices. Representing the unit vectors in spherical coordinates, we have, $\vec{x} = (\sin\theta a_0 \cos\phi a_0, \sin\theta a_0 \sin\phi a_0, \cos\theta a_0), ~~\vec{y} = (\sin\alpha b_0 \cos\beta b_0, \sin\alpha b_0 \sin\beta b_0, \cos\alpha b_0) $ and $\vec{z} = (\sin\zeta c_0 \cos\eta c_0, \sin\zeta c_0 \sin\eta c_0, \cos\zeta c_0) $ and similarly, we define, $\vec{\acute{x}},\vec{\acute{y}}$ and $\vec{\acute{z}}$ by replacing $0$ in the indices by $1$. Then the value of the operator $S$ (Eq.(\ref{2i})) with respect to the state $\rho_1$ (Eq.(\ref{1})) gives:
\begin{widetext}
	$S(\rho_1) = \cos(\alpha b_0) (-1+p_1+ p_1\cos(2 \theta_1))(\cos(\zeta c_0)(\cos( \theta a_0)-\cos(\theta a_1))+\cos(\zeta c_0)(\cos\theta a_0)+ \cos(\theta a_1))+\cos(\alpha b_1)(-1+p_1+ p_1\cos(2 \theta_1))(\cos(\zeta c_0)(\cos(\theta a_0)-\cos(\theta a_1))-\cos(\zeta c_1)(\cos(\theta a_0)+ \cos(\theta a_1)))+ p_1\sin(2\theta_1)(\cos(\beta b_0+\eta c_0+\phi a_0)\sin(\alpha b_0)\sin(\zeta c_0)\sin(\theta a_0)+\cos(\beta b_1+\eta c_0+\phi a_0) \sin(\alpha b_1)\sin(\zeta c_0)\sin(\theta a_0)+\cos(\beta b_0+\eta c_1+\phi a_0)\sin(\alpha b_0)\sin(\zeta c_1)\sin(\theta a_0)-\cos(\beta b_1+\eta c_1+\phi a_0)\sin(\alpha b_1)\sin(\zeta c_1)\sin(\theta a_0)+\cos(\beta b_0+\eta c_0+\phi a_1)\sin(\alpha b_0)\sin(\zeta c_0)\sin(\theta a_1)-\cos(\beta b_1+\eta c_0 +\phi a_1)\sin(\alpha b_1)\sin(\zeta c_0)\sin(\theta a_1)$
	\begin{equation}\label{B1}
	- \cos(\beta b_0+\eta c_1+\phi a_1)\sin(\alpha b_0)\sin(\zeta c_1)\sin(\theta a_1)-\cos(\beta b_1+\eta c_1+\phi a_1)\sin(\alpha b_1)\sin(\zeta c_1) \sin(\theta a_1)).
	\end{equation}
\end{widetext}
Hence in order to get maximum value of $S(\rho_1)$, we have to perform a maximization over $12$ measurement angles. We first find the global maximum of $S(\rho_1)$ with respect to $\theta a_0$ and $\theta a_1$. We begin with by finding all critical points of $S(\rho_1)$ inside the region $R=[0 , 2\pi]\times [0 , 2\pi]$ which are namely $(0,0)$, $(\frac{\pi}{2}, -\frac{\pi}{2})$,$(-\frac{\pi}{2}, \frac{\pi}{2})$ , $(\frac{\pi}{2}, \frac{\pi}{2})$ and $(-\frac{\pi}{2}, -\frac{\pi}{2})$. Among all these critical points, $(\frac{\pi}{2}, \frac{\pi}{2})$ gives the global maximum of $S(\rho_1)$ if $\sqrt{2} p_1 \sin(2 \theta_1) > |(1- p_1 - p_1 \cos(2 \theta_1) )|$. Thus, we have:
\begin{widetext}
	$ S(\rho_1)\leq p_1\sin(2\theta_1)(\cos(\beta b_0+\eta c_0+\phi a_0)\sin(\alpha b_0)\sin(\zeta c_0)\sin(\theta a_0)+\cos(\beta b_1+\eta c_0+\phi a_0)\sin(\alpha b_1)\sin(\zeta c_0)\sin(\theta a_0)+\cos(\beta b_0 + \eta c_1 + \phi a_0) \sin(\alpha b_0) \sin(\zeta c_1) \sin(\theta a_0) - \cos(\beta b_1 + \eta c_1 + \phi a_0) \sin(\alpha b_1] \sin(\zeta c_1) \sin(\theta a_0) + \cos(\beta b_0 + \eta c_0 + \phi a_1) \sin(\alpha b_0) \sin(\zeta c_0) \sin(\theta a_1)-\cos(\beta b_1+\eta c_0+\phi a_1)\sin(\alpha b_1)\sin(\zeta c_0)\sin(\theta a_1) $
	\begin{equation}\label{B2}
	-\cos(\beta b_0+\eta c_1+\phi a_1)\sin(\alpha b_0) \sin(\zeta c_1)\sin(\theta a_1)-\cos(\beta b_1+\eta c_1+\phi a_1)\sin(\alpha b_1)\sin(\zeta c_1)\sin(\theta a_1)).
	\end{equation}
\end{widetext}
Now we carry out the same procedure over the following pair of variables $(\zeta c_0, \zeta c_1)$ and $(\alpha b_0, \alpha b_1)$, one by one. Similar to the previous case, critical point $(\frac{\pi}{2}, \frac{\pi}{2})$ gives the maximum value for both of these pair of variables. So, the last inequality in Eq.(\ref{B2}) takes the form
\begin{widetext}
	$S(\rho_1) \leq  p_1 \sin(2 \theta_1)(\cos \eta c_0 (\cos(\beta b_0 + \phi a_0)+\cos(\beta b_1 + \phi a_0)+\cos(\beta b_0 + \phi a_1)-\cos(\beta b_1 + \phi a_1))-\sin \eta c_0(\sin(\beta b_0 + \phi a_0)+\sin(\beta b_1 + \phi a_0)+\sin(\beta b_0 + \phi a_1)-\sin(\beta b_1 + \phi a_1))+\cos \eta c_1 (\cos(\beta b_0 + \phi a_0)-\cos(\beta b_1 + \phi a_0)$
	\begin{equation}\label{B4}
	-\cos(\beta b_0 + \phi a_1)-\cos(\beta b_1 + \phi a_1))+\sin \eta c_1(-\sin(\beta b_0 + \phi a_0)+\sin(\beta b_1 + \phi a_0)+\sin(\beta b_0 + \phi a_1)+\sin(\beta b_1 + \phi a_1))).
	\end{equation}
\end{widetext}
which when maximized with respect to $\eta c_0$ and $\eta c_1$ gives:\\
\begin{widetext}
\begin{equation}\label{B5i}
S(\rho_1) \leq 2 p_1 \sin(2 \theta_1)\sqrt{(\cos A_{00}+\cos A_{10}+\cos(A_{01})-\cos A_{11})^2+(\sin A_{00}+\sin A_{10}+\sin A_{01}-\sin A_{11})^2}
\end{equation}
\end{widetext}
where $A_{ij}=\beta b_i + \phi a_j,\,(i,j\in\{0,1\})$.
The last inequality is obtained by using the inequality $x\cos\theta + y \sin\theta \leq \sqrt{x^2 + y^2}$. Maximum value of the expression in Eq.(\ref{B5i}) remains unaltered by putting any value of $\beta b_0$ and $\phi a_0$. In particular if we take $\beta b_0 = 0$ and $\phi a_0=0$, then maximum value is obtained for $(\beta b_1, \phi a_1 ) = (\frac{\pi}{2}, \frac{\pi}{2})$ and is equal to $ 4\sqrt{2} p_1 \sin 2\theta_1 $.
Again if $\sqrt{2} p_1 \sin(2 \theta_1) < |(1- p_1 - p_1 \cos(2 \theta_1) )|$, the critical point $(0, 0)$ gives the maximum value of $S(\rho_1)$. Then Eq.(\ref{B1}) reduces to $S(\rho_1) \leq 2(-1 + p_1 + p_1 \cos(2 \theta_1))(\cos(\alpha b_0)\cos(\zeta c_0)-\cos(\alpha b_1)\cos(\zeta c_1)).$
Now the critical point $(0,0)$ gives the maximum value when we maximize the last expression with respect to $\alpha b_0$ and $\alpha b_1 $ . Then the last inequality becomes $S(\rho_1) \leq 2(-1 + p_1 + p_1 \cos(2 \theta_1))(\cos(\zeta c_0) - \cos(\zeta c_1)).$
Again we maximize it with respect to $\zeta c_0$ and $\zeta c_1$. Critical point $(0,\pi)$ or $(\pi,0)$ gives the maximum value depending on whether  $p_1(1 +  \cos(2 \theta_1))> 1$ or $p_1(1 +  \cos(2 \theta_1))< 1$. Hence in any case $S(\rho_1) \leq 4|1- p_1 - p_1 \cos(2 \theta_1)|$. So $S(\rho_1) \leq \max[4\sqrt{2} p_1 \sin 2\theta_1 , 4|1- p_1 - p_1 \cos(2 \theta_1)|]$ as stated in Eq.(\ref{6i}) of main text. Similarly one can obtain $B_i$ $(i= 2,3,4)$. From these values of $B_i$(i = 1, 2, 3), one can get the range of $p_i$ for which the corresponding initial state $\rho_i$ satisfy Svetlichny inequality. For the final state $\rho_4$, we obtain the range of violation of Svetlichny inequality, by following the above analytical method. We now proceed to search for the maximum expectation value of operator($NS_i(\rho_j),j=1,2,3,4$) corresponding to the remaining $i^{th},(i=1,2,...,184)$ facet inequality.  \\

\textit{Checking the remaining $184$ facets:} The above method of maximization is applied for most of the remaining $184$ facet inequalities excluding a few for which the upper bound of violation($B_i(i=1,2,3)$) is measurement specific, i.e. varies not only with the state parameters but also with variation of parameters characterizing the measurement settings. In order to find the range of $p_i$ for those inequalities, we have performed numerical optimization by using Mathematica software \cite{Wol}.
We now give an example of such a facet inequality for which the analytical method of maximization does not hold good due to dependence of the upper bound of expectation value of the corresponding operator over measurement settings apart from state parameters. $3^{rd}$ facet(say), is  given by : $NS_3$ =
$$-\langle x_0\rangle-\langle x_1\rangle-\langle x_0y_0\rangle -2\langle y_1 \rangle  -\langle z_0\rangle+\langle x_1y_0\rangle-\langle x_0z_0\rangle$$
$$+\langle y_0z_0\rangle + \langle x_1y_0z_0\rangle-\langle x_0y_1z_0\rangle+\langle x_1y_1z_0\rangle-\langle z_1\rangle+$$
\begin{equation}\label{B55}
\langle x_1z_1\rangle-\langle y_0z_1\rangle-\langle x_0y_0z_1\rangle+\langle x_0y_1z_1\rangle+\langle x_1y_1z_1\rangle \leq 4.
\end{equation}
The value of the operator $NS_3$ given by the $3^{rd}$ facet with respect to the state $\rho_1$(Eq.(\ref{1}) of main text) under the projective measurement gives:
\begin{widetext}
	$$NS_3(\rho_1)= \cos\zeta c_1((1-p_1)(1+\cos\alpha b_0+\cos\theta a_0\cos\alpha b_0 )-p_1 \cos^2 \theta_1-p_1\cos\alpha b_0\cos^2 \theta_1)-(1-p_1)\cos\theta a_0( 1+\cos\alpha b_0)$$
	$$-p_1\cos^2\theta_1\cos\theta a_0(1+\cos\alpha b_0+\cos\alpha b_0\cos\zeta c_1)+(\cos\alpha b_0 \cos\theta a_1-\cos\zeta c_1\cos\theta a_1-1)(1-p_1)-p_1 \cos^2 \theta_1 \cos\theta a_1(1-\cos\alpha b_0$$
	$$ -\cos\zeta c_1)+ \cos\zeta c_0(1-p_1-p_1\cos 2\theta_1+\cos\theta a_0(1-2p_1)+\cos\alpha b_0(-1+2p_1+(-1+p_1+p_1\cos2\theta_1)\cos \theta a_1))$$
	$$-\frac{1}{2}\cos\alpha b_1(\cos\zeta c_0(-2+p_1+p_1\cos2\theta_1)(\cos\theta a_0-\cos\theta a_1)-2p_1\cos^2\theta_1(-2+\cos\zeta c_0(\cos \theta a_1-\cos \theta a_0)+\cos\zeta c_1(\cos \theta a_1$$
	$$+\cos \theta a_0))+2(2-3p_1+p_1\cos2\theta_1-\frac{1}{2}\cos\zeta c_1(-2+p_1+p_1\cos2\theta_1)(\cos \theta a_1+\cos \theta a_0))+(p_1\cos\zeta c_1\sin^2\theta_1$$
	$$+p_1\cos\theta a_0\sin^2\theta_1)(1-\cos\alpha b_0)+p_1\cos\alpha b_0\cos\zeta c_1\cos\theta a_0\sin^2\theta_1+p_1\cos\theta a_1\sin^2\theta_1(1+\cos\alpha b_0+\cos\zeta c_1)-$$
	$$p_1\cos(\beta b_1+\eta c_0+\phi a_0)\sin\alpha b_1\sin\zeta c_0\sin 2\theta_1\sin\theta a_0-p_1\cos(\beta b_0+\eta c_1+\phi a_0)\sin\alpha b_0\sin\zeta c_1\sin 2\theta_1\sin\theta a_0+$$
	$$p_1\cos(\beta b_1+\eta c_1+\phi a_0)\sin\alpha b_1\sin\zeta c_1\sin 2\theta_1\sin\theta a_0+p_1\cos(\beta b_0+\eta c_0+\phi a_1)\sin\alpha b_0\sin\zeta c_0\sin 2\theta_1\sin\theta a_1$$
	\begin{equation}\label{B56}
	+p_1\cos(\beta b_1+\eta c_0+\phi a_1)\sin\alpha b_1\sin\zeta c_0\sin 2\theta_1\sin\theta a_1+p_1\cos(\beta b_1+\eta c_1+\phi a_1)\sin\alpha b_1\sin\zeta c_1\sin 2\theta_1\sin\theta a_1)
	\end{equation}
\end{widetext}
Now to find the upper bound of $NS_3(\rho_1)$ in terms of state parameters, we need to maximize $NS_3(\rho_1)$ over all the variables parameterizing measurement settings. However, for almost each of those variables there is no fixed critical point for which $NS_3(\rho_1)$ gives maximum value, it varies with the variation of state parameters. Hence, the analytical method that was followed for $S(\rho_1)$ cannot be applied. In order to overcome this difficulty, we apply numerical optimization by using Mathematica Software \cite{Wol}. We consider a particular example. Let $\theta_1 = 0.1$. The measurement settings parameters vary with the other state parameter $p_1,$ i.e., the maxima of $NS_3(\rho_1)$ with respect to any measurement parameter varies with state parameter $p_1.$  So we maximize $NS_3(\rho_1)$ over all measurement parameters by using Mathematica Software. After maximizing numerically, it is observed that under the restriction $0\leq p_1 \leq 0.509$, the maximum value of $NS_3(\rho_1)$ never exceeds $4$. Hence the initial state $\rho_1$ with $\theta_1 = 0.1$ satisfies $3-$rd facet when $0\leq p_1 \leq 0.509$. We have applied this numerical method for all the facets for which the upper bound of violation depends over measurement settings apart from state parameters. In totality, i.e.  considering all facets(some by analytical method and others by numerical method), it is checked that $\rho_1$ with $\theta_1 = 0.1$ satisfy all of the $185$ facets when $0\leq p_1 \leq 0.509$. Similar method is applied to find the range of $p_1$  for which $\rho_1$ satisfy all of $185$ facets for different fixed values of $\theta_1$. Just as for the initial state $\rho_1$, we have followed similar trend of analysis for the other two initial states $\rho_2$, $\rho_3$ and also for the resultant state $\rho_4$.

\subsection{Local filtering and hidden Genuine Tripartite nonlocality}
Here we will discuss the effect of using local filtering on the initial states $\rho_i(i=1,2,3)$. Any local filtering transforms a tripartite state $\rho$ in
\begin{equation}\label{B50}
\acute{\rho} = \frac{(F_1 \bigotimes F_2 \bigotimes F_3) \rho (F^{\dag}_1 \bigotimes F^{\dag}_2 \bigotimes F^{\dag}_3)}{\textmd{tr}((F_1 \bigotimes F_2 \bigotimes F_3) \rho (F^{\dag}_1 \bigotimes F^{\dag}_2 \bigotimes F^{\dag}_3))}
\end{equation}
where $F^{\dag}_j F_j \leq I_2 $ $(j=1, 2, 3)$. It is shown in \cite{Hic} for qubit case, the most general filters are of the form $F_j$ =
\[ \left( \begin{array}{cc}
\epsilon_j & 0\\
0 & 1
\end{array}\right)
\left( \begin{array}{cc}
\cos\theta_j & -e^{i\phi_j \sin\theta_j}\\
e^{i\phi_j \sin\theta_j} & \cos\theta_j
\end{array}\right)
\]
where $\epsilon_j$, $\theta_j$, $\phi_j$ are real parameters.  It is argued in \cite{Hic} that theoretically there is no reason to exclude the unitary matrix
in $F_j$ (which corresponds to a local unitary before the filter), yet in standard form of local filters, the contribution from the unitary matrix is ignored. In \cite{Hic} it is also argued that all the known useful filters are diagonal. Especially for the qubit case, it seems that only the diagonal filters are relevant. Since we are dealing with qubit case only, we take the filters of the form $F_j$ =
\[ \left( \begin{array}{cc}
\epsilon_j & 0\\
0 & 1
\end{array}\right).
\]
Here $ \epsilon_j$'s($j=1,2,3$) are filtering parameters and $0 \leq \epsilon_j \leq 1$.
Now the application of local filtering on the state $\rho_1$(Eq.(\ref{1}) in the main text) results in
\begin{equation}\label{B51}
\acute{\rho_1} =  \frac{p_1 |\phi_1\rangle\langle \phi_1| + (1 - p_1)\epsilon^2_2 \epsilon^2_3 |100\rangle\langle100|}{(1-p_1)\epsilon^2_2 \epsilon^2_3 + p_1 \epsilon^2_1 \epsilon^2_2 \epsilon^2_3 \cos^2\theta_1 + p_1 \sin^2\theta_1}.
\end{equation}
where $|\phi_1\rangle = \epsilon_1 \epsilon_2 \epsilon_3 \cos\theta_1 |000\rangle + \sin\theta_1 |111\rangle.$\\
To obtain the maximum value of the Svetlichny operator $S$ with respect to projective measurements, for the state $\acute{\rho_1}$, we apply the same method as we used in the last section for the derivation of $B_1$. The maximum value is given by
$$\max[\frac{4\sqrt{2}p_1\epsilon_1 \epsilon_2 \epsilon_3 \sin2\theta_1}{(1-p_1)\epsilon^2_2 \epsilon^2_3 + p_1 \epsilon^2_1 \epsilon^2_2 \epsilon^2_3 \cos^2\theta_1 + p_1 \sin^2\theta_1},$$
\begin{equation}\label{B52}
\frac{4((1-p_1)\epsilon^2_2 \epsilon^2_3 - p_1 \epsilon^2_1 \epsilon^2_2 \epsilon^2_3 \cos^2\theta_1 + p_1 \sin^2\theta_1)}{(1-p_1)\epsilon^2_2 \epsilon^2_3 + p_1 \epsilon^2_1 \epsilon^2_2 \epsilon^2_3 \cos^2\theta_1 + p_1 \sin^2\theta_1}].
\end{equation}
Clearly,$\frac{4((1-p_1)\epsilon^2_2 \epsilon^2_3 - p_1 \epsilon^2_1 \epsilon^2_2 \epsilon^2_3 \cos^2\theta_1 + p_1 \sin^2\theta_1)}{(1-p_1)\epsilon^2_2 \epsilon^2_3 + p_1 \epsilon^2_1 \epsilon^2_2 \epsilon^2_3 \cos^2\theta_1 + p_1 \sin^2\theta_1} \leq 4$ for any value of $0\leq p_1 \leq 1$ and $0 \leq \epsilon_j \leq 1$. So the filtered state $\acute{\rho_1}$ remains $S_2$ local if $\frac{4\sqrt{2}p_1\epsilon_1 \epsilon_2 \epsilon_3 \sin2\theta_1}{(1-p_1)\epsilon^2_2 \epsilon^2_3 + p_1 \epsilon^2_1 \epsilon^2_2 \epsilon^2_3 \cos^2\theta_1 + p_1 \sin^2\theta_1} \leq 4$. After maximizing the left hand side of the last inequality with respect to $\epsilon_j$ $(j=1,2,3)$,  we have
\begin{equation}\label{B55}
p_1 \leq \frac{2}{3+\cos2\theta_1}
\end{equation}
From Eq.(\ref{B55}) one can get the range of $p_1$ for each non-zero value of $\theta_1$ such that the filtered state $\acute{\rho_1}$ remains $S_2$ local, i.e. the initial state $\rho_1$ has no hidden $S_2$ nonlocality. Similarly the range of $p_1$ for each  non-zero value of $\theta_1$ for which the filtered state $\acute{\rho_1}$ satisfies remaining facet inequalities are obtained. For most of the facet inequalities, the analytical method(as followed in the previous section) is applicable excepting a few where the upper bound of the expectation value of the operator $NS_i(\rho_1)$ corresponding to the $i^{th}(i=1,...,184)$ facet depends not only on the state parameters but also on the variables parameterizing measurement settings. For those few facets we have done numerical optimization by Mathematica software(as already discussed in the previous section).
For instance, we consider $3^{rd}$ facet inequality. Let us fix the state parameter $\theta_1$: $\theta_1=0.1$. For this fixed value of $\theta_1$, numerical maximization of $NS_3(\rho_1)$ over all the measurement settings shows that under the restriction $\epsilon_j(j=1,2,3)\in[0,1]$ and $p_1\in[0,0.515]$, state $\acute{\rho_1}$ satisfies $3^{rd}$ facet inequality. After checking all of $NS_i(\rho_1),i=1,...,185$, we arrive at the conclusion that for $\theta_1=0.1$ and $p_1\in[0,0.5025]$, state $\rho_1$ does not reveal any GTNL after the application of known useful local filters.  We have applied the same procedure over other fixed values of $\theta_1$. For other two initial states $\rho_2$ and $\rho_3$, we have made analysis in similar manner so as to obtain the range of $p_2$ and $p_3$(for a fixed value of $\theta_3$) of $\rho_2$(Eq.(\ref{2}) of the main text) and $\rho_3$(Eq.(\ref{3}) of the main text) respectively for which they still do not reveal any hidden GTNL after the application of local filters.

\end{document}